# Demonstration of Time-reversal Symmetric Two-Dimensional Photonic Topological Anderson Insulator

Zhe Li[1], Ziming Chen[1], Deyang Kong[1], Yongzhuo Li[1†], Kaiyu Cui[1], Xue Feng[1‡], Yidong Huang[1§]

[1]Department of Electronic Engineering, Tsinghua University, 100084 Beijing, China.

Recently, the impact of disorder on topological properties has attracted significant attention in photonics, especially the intriguing disorder-induced topological phase transitions in photonic topological Anderson insulators (PTAIs). However, the reported PTAIs are based on time-reversal symmetry broken systems or quasi-three-dimensional time-reversal invariant system, both of which would limit the applications in integrated optics. Here, we realize a time-reversal symmetric two-dimensional PTAI on silicon platform within the near-IR wavelength range, taking the advantageous valley degree of freedom of photonic crystal. A low-threshold topological Anderson phase transition is observed by applying disorder to the critical topologically trivial phase. Conversely, we have also realized extremely robust topologically protected edge states based on the stable topological phase. Both two phenomena are validated through theoretical Dirac Hamiltonian analysis, numerical simulations, and experimental measurements. Our proposed structure holds promise to achieve near-zero topological phase transition thresholds, which breaks the conventional cognition that strong disorder is required to induce the phase transition. It significantly alleviates the difficulty of manipulating disorder and could be extended to other systems, such as condensed matter systems where strong disorder is hard to implement. This work is also beneficial to construct highly robust photonic integrated circuits serving for on-chip photonic and quantum optic information processing. Moreover, this work also provides an outstanding platform to investigate on-chip integrated disordered systems.

Photonic topological insulators (PTIs), which possess topological band gaps and robust edge states against weak disorder [1–3], have gained much attention for the potential applications in photonic integrated circuits (PICs) and lasers [4–12]. Previous reports indicate that disorder is detrimental to PTIs since extremely strong disorder would lead the system to topologically trivial [13–15], as well as localized states due to Anderson localization [16–18].

However, counter-intuitively, strong disorder could also introduce nontrivial topology into a topologically trivial system, which was first proposed in condensed-matter system by Li *et al* and is so called topological Anderson insulators [19,20]. In succession, photonic topological Anderson insulators (PTAIs) have also been demonstrated in several systems [21–26], including quasi-three-dimensional (quasi-3D) and two-dimensional (2D) photonic systems. For the quasi-3D systems, both time-reversal symmetry broken [21] and time-reversal symmetric [25] PTAIs have been achieved, while for 2D system, only time-reverse symmetry broken systems with gyromagnetic materials and corresponding external magnetic fields have been realized [22]. Obviously, both of such 3D structure and external magnetic field (or time-varying gauge field) needed in 2D system are not feasible for the on-chip photonic integration. Thus, a time-reversal symmetric 2D PTAI is still desired to achieve an ultra-compact and robust photonic integrated circuit as well as investigate on-chip integrated disordered system.

Here, we experimentally demonstrate a time-reversal symmetric 2D PTAI on the all-dielectric silicon (Si) platform employing the valley photonic crystals (VPCs) with two opposite valley Chern numbers by controlling the oriental angle of the atoms inside the unit cell. Disorder is introduced by randomly rotating the atoms in each unit cell. The VPCs gradually change from topologically trivial into topologically non-trivial systems with the increasing disorder strength, and thus a valley topologically protected Anderson photonic crystal (VTAPC) is obtained. In our proposed scheme, slight disorder would be enough to induce the topological phase transition, indicating a low phase transition threshold. It also means that our proposed structure could achieve extremely robust topologically protected edge states against strong disorder (maximum random rotation angle of 60°). What's more, different from most of previous works operating within the microwave range, our demonstrated VTAPC operates in the near-IR wavelength and is also, to the best of our knowledge, the first 2D integrated PTAI, which would provide a flexible platform to investigate on-chip disordered topological systems.

[†]Yongzhuo Li: liyongzhuo@tsinghua.edu.cn

[‡]Xue Feng: x-feng@tsinghua.edu.cn

[§]Yidong Huang: yidonghuang@tsinghua.edu.cn

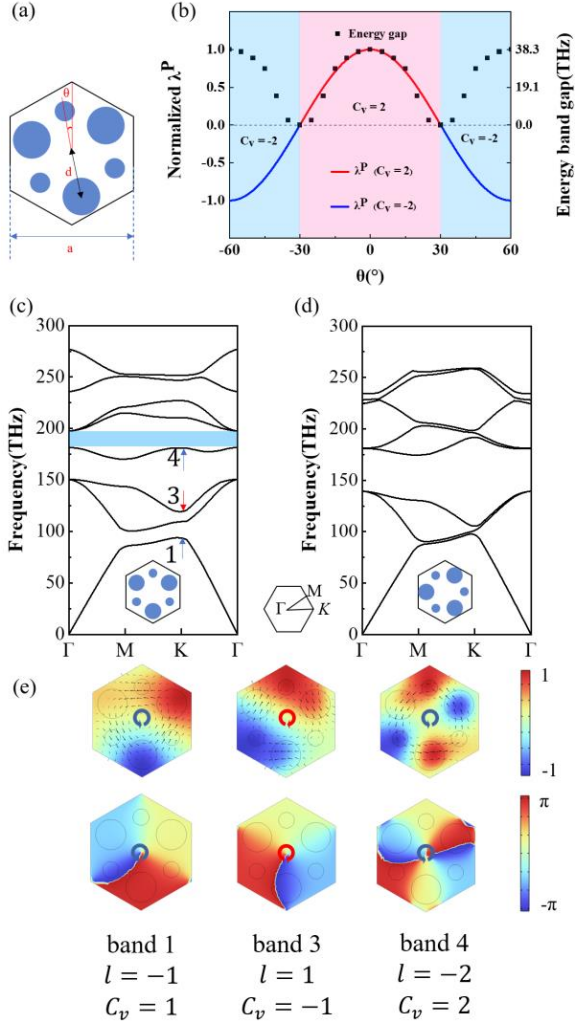

FIG. 1. Phase diagram and energy band structure of the valley photonic crystal. (a) Scheme of the unit cell of the VPC. (b) Phase diagram of the VPC, the black solid squares are valley energy gaps; blue (red) lines are normalized $\lambda^P$ with valley Chern number -2 (2); while red (blue) shallow regions represent the topological phase with valley Chern number -2 (2). (c) Band structure of the VPC with $\theta = 0°$, a band gap with $C_v = 2$ is marked by blue. At $K$ valley, three modes (marked by arrows) are with non-zero valley Chern number. (d) Band structure of the VPC with $\theta = 30°$. (e) Mode profiles of the three modes marked in (c). Upper: $E_z$ field distributions and Poynting energy flows (black arrows), the red (blue) circular arrows mark clockwise (anticlockwise) Poynting energy flows. Below: Phases of $E_z$, the red (blue) circular arrows mark positive (negative) OAM charge, which is represented by $l$.

†Yongzhuo Li: liyongzhuo@tsinghua.edu.cn

‡Xue Feng: x-feng@tsinghua.edu.cn

§Yidong Huang: yidonghuang@tsinghua.edu.cn

*Design of time-reversal symmetric 2D PTAI—* In this work, our proposed 2D PTAI structures are prepared on 1000-nm-thickness silicon (Si) layer. As shown in Fig. 1(a), the $C_3$ −symmetric honeycomb-shaped unit cell of the VPC is composed of six Si rods with two different radii $R_A$ and $R_B$ surrounded by air, and the rods could be rotated around the center of the unit cell. The structural parameters are as lattice constant $a = 900$ nm, height $h = 1000$ nm, $R_A = 0.15a$, $R_B = 0.08a$, the distance between the centers of rods and the unit cell $d$, and the oriental angle of the rods $\theta$. Due to $C_3$ symmetry of the unit cell, the tuning range of $\theta$ could be from -60° to 60°. Such parameters are meticulously selected to operate within the near-IR wavelength range, considering for the application of photonic integration. With such structure, the energy bands around $K$ and $K'$ valleys can be described by the Dirac Hamiltonian:

$$\hat{H} = f_D + \upsilon_D \left( \hat{\sigma}_x \hat{\tau}_z \delta k_x + \hat{\sigma}_y \delta k_y \right) + \lambda^P \hat{\sigma}_z, \quad (1)$$

$$\lambda^P \propto 2\left( \int_A \varepsilon(\vec{r})dS - \int_B \varepsilon(\vec{r})dS \right) \times \\ \left[ \sin\left(\vec{G_1} \cdot \vec{r_{A_1}}\right) + \sin\left(\vec{G_1} \cdot \vec{r_{A_2}}\right) + \sin\left(\vec{G_1} \cdot \vec{r_{A_3}}\right) \right], \quad (2)$$

where $\hat{\sigma}_i$, $\hat{\tau}_i$ are the Pauli matrices acting on orbit and valley, and $\lambda^P$ is the effective mass term, $|\lambda^P|$ and $\text{sgn}(\lambda^P)$ describes energy gap and valley Chern number ($C_v$) at $K/K'$ valley, respectively. In Eq. (2), $\vec{G_1}$ is a reciprocal vector of the unit cell, $\vec{r_{A_i}}$ is the position of the rods with radius $R_A$, and thus $\lambda^P$ is determined by $\vec{r_{A_i}}$, which is determined by $d$ and $\theta$ (details shown in Supplemental Material. A) [27–29]. For a fixed $\theta$, $d$ determines the $|\lambda^P|$, so that we set $d = 0.6a/\sqrt{3}$ to achieve the largest energy gap. The phase diagram with respect to $\theta$ is shown in Fig. 1(b), the solid lines represent the angle-dependent normalized $\lambda^P$ values calculated by Eq. (2) (blue and red colors represent valley Chern numbers of -2 and 2, respectively), and the black solid squares show the valley energy gap obtained from finite element method simulations. The evolution of $|\lambda^P|$ versus $\theta$ is generally consistent with that of the energy gap.

The phase diagram could be divided to two regions: when $|\theta| < 30°$, $\lambda^P > 0$, the corresponding valley Chern number is $C_v = 2$ (details are shown later); when $|\theta| > 30°$, $\lambda^P < 0$, $C_v = -2$. Only when $\theta = \pm 30°$, $\lambda^P = 0$ and $C_v = 0$, which is the topologically trivial case. Actually, the mass term $\lambda^P$ represents the parity symmetry breaking (PSB) strength of the system, which would lead to valley degeneration breaking and energy gap opening. When $\theta = \pm 30°$, $\lambda^P = 0$, PSB would achieve the minimum value (parity symmetric), and the $K$ and $K'$ valleys are degenerate and topologically trivial; while PSB and corresponding bandgap would achieve the maximum value at $\theta = 0°, \pm 60°$. Fig. 1(c) shows the energy bands of the transverse magnetic (TM) modes for the case of $\theta = 0°$. There are four bands below the band gap (blue region), three of them with non-zero valley Chern number. The corresponding intensity and phase profiles of $E_z$ at $K$ point are shown in Fig. 1(e). The energy flow directions and orbital angular momentum (OAM) charges indicate the valley Chern numbers [30] (Details in Supplemental Material. B). Therefore, the $C_v$ of the energy gap is 2 (sum of valley Chern numbers of all bands below the gap). For comparison, Fig1. (d) shows another case of $\theta = 30°$, in which energy gap is closed. Obviously, it is topologically trivial.

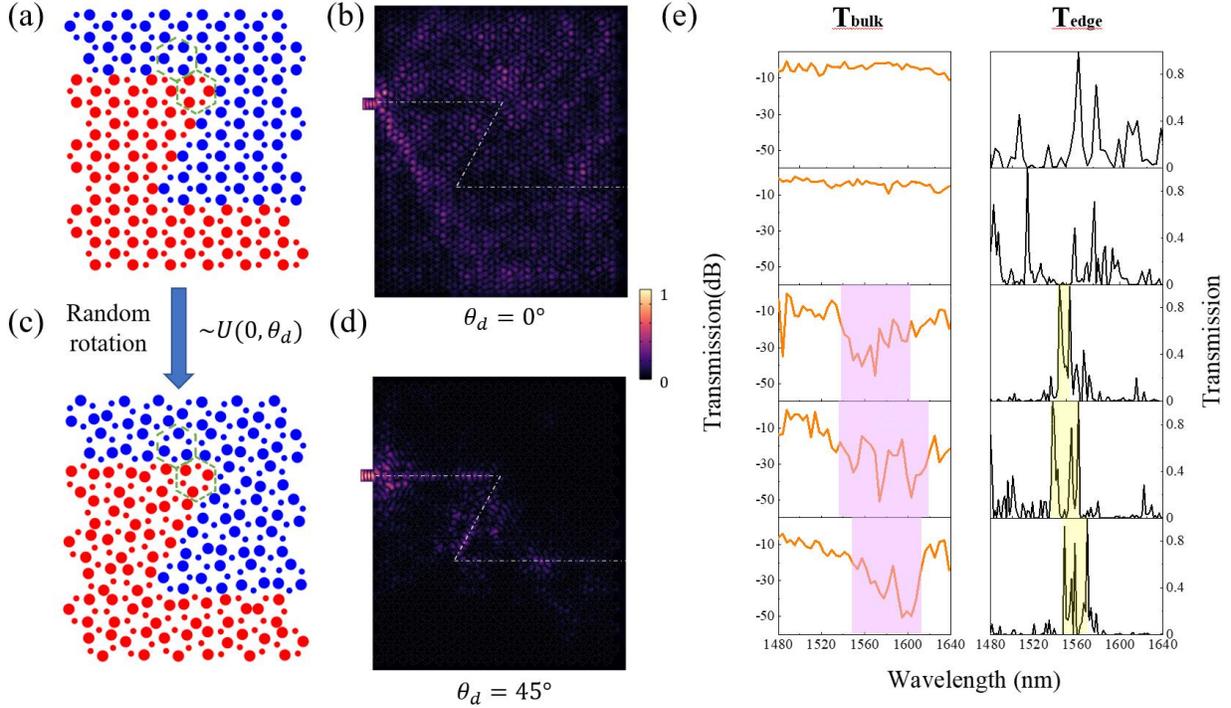

Fig. 2. Schematic of the VTAPC. (a) Schematic of the critical topologically trivial photonic crystal composed of two VPCs with $\theta = \pm 30°$ (blue for 30°, red for −30°). (b) Simulated field intensity of $|E_z|$ for $\theta_d = 0°$ at 1555 nm, the interface is marked by white dashed line. (c) Schematic of VTAPC, disorder is introduced by randomly rotating the rods in each unit cell. (d) Simulated field intensity of $|E_z|$ of the VTAPC with $\theta_d = 45°$, the interface is marked by white dashed line. (e) Simulated bulky and edge transmission spectra with $\theta_d = 0°$, 15°, 30°, 45° and 60°, from top to bottom. Bulk energy gap is marked by purple rectangular (defined by transmission < -20 dB), while edge transmission width is marked by yellow rectangular. All the spectra are normalized.

In the parameter space of $\theta$, only two points with $\theta = \pm 30°$ have zero mass term $\lambda^P = 0$ and are corresponding to critical topologically trivial phase, so that slight variation of $\theta$ could induce a transition from the trivial phase to the non-trivial phase. Then, a critical topological trivial system could be achieved by VPCs with $\theta = \pm 30°$, as shown in Fig. 2(a). The two VPCs are arranged to form a Z-shaped interface to investigate the topological protection properties. In this system, since valley Chern numbers of two VPCs are both zero, there is no bulky energy gap and topologically protected edge mode (Details in Supplemental Material. C). Therefore, the bulk mode light-wave transmits inside the photonic crystal, and the transmission profile is plotted in Fig. 2(b). Disorder is introduced by randomly anticlockwise rotating the Si rods in each unit cell (as shown in Fig. 2(c)), the random rotation angle follows the uniform distribution of $\theta_i \sim U(0, \theta_d)$. The random rotation angle range $\theta_d$ is used to represent the strength of the disorder. With disorder, the $C_v$ of the two VPCs would have opposite sign ($C_v = -2, 2$ respectively), the bulky energy gap would open while the valley topologically protected edge modes emerge simultaneously. In order to maintain the opposite $C_v$ of two disordered VPCs, the maximum possible value of $\theta_d$ is 60°.

†Yongzhuo Li: liyongzhuo@tsinghua.edu.cn

‡Xue Feng: x-feng@tsinghua.edu.cn

§Yidong Huang: yidonghuang@tsinghua.edu.cn

To provide a theoretical analysis, the effective Hamiltonian of the disordered VPC could be formulated by averaging the $\lambda^P$ of each unit cell:

$$\hat{H} = f_D + v_D\left(\hat{\sigma}_x \hat{\tau}_z \delta k_x + \hat{\sigma}_y \delta k_y\right) + \sum_{i=1}^{N} \frac{\lambda_i^P(\theta_i)}{N} \hat{\sigma}_z, \quad (3)$$

where $\lambda_i^P$ is the $\lambda^P$ within each unit cell and determined by the corresponding rotation angle. The averaged efficient mass term is given by $\lambda_e^P = \sum_{i=1}^{N} \frac{\lambda_i^P(\theta_i)}{N}$ [25]. It should be noted that $|\lambda_e^P|$ represents the bandgap at the valley, not the complete bandgap of the system, which is actually smaller than the former. This implies that when $\lambda_e^P$ is small, a complete bandgap does not exist. As $\theta_d$ increasing, PSB firstly increases and reaches the maximum value at 45°, then would decrease. With strong enough PSB, the energy gap would open and the disordered photonic crystal would undergo the topological phase transition, turn into VTAPC and support topologically protected edge modes.

Furthermore, we have simulated the bulk and edge transmission of VTAPCs with $\theta_d = 0°$, 15°, 30°, 45° and 60° (details in Supplemental Material. D), respectively. Fig. 2(d) shows an edge transmission case of $\theta_d = 45°$, a topologically protected edge mode emerging around the interface. The corresponding transmission spectra are also calculated and shown in Fig. 2(e). When $\theta_d = 0°$ and 15°, there is no bulky energy gap and edge transmission at the interface of the VPCs due to relatively weak PSB. However, with stronger disorder ($\theta_d \geq 30°$), the bulky energy gap would open and edge transmission would emerge. Specifically, the maximum width is achieved with $\theta_d = 45°$. These results indicate that the topological phase transition has already occurred at $\theta_d = 30°$.

Topological Anderson phase transitions reported in previous works [21,25] typically start from a topologically trivial system with trivial energy gap, experience gradual closing and reopening of the bulk energy gap, in which strong enough disorder is required to induce the transition. As comparison, our approach starts from critical topologically trivial phases with closed bulk energy gap and the gap gradually opens with increasing disorder strength. As long as a complete band gap is achieved, topologically protected waveguide would emerge, resulting a much lower phase transition threshold.

*Extremely Robust Valley topologically protected photonic crystal against strong disorder*—In the previous section, it has been demonstrated that a small amount of disorder could induce a topological phase transition. Conversely, it also implies that strong disorder would not be easy to turn a valley topological insulator into trivial phase. In the phase diagram (Fig. 1(b)), the extreme value points ($\theta = 0°, \pm 60°$) are corresponding to the stable topological phase, as its valley Chern number would not change even under strong disorder. An extremely robust valley topologically protected photonic crystal (VTPC) is composed of two VPCs with stable topological phases ($\theta = 0°, 60°$), as shown in Fig. 3(a), the interface is also Z-shaped. Such two VPCs have opposite $C_v$ and thus could support valley topologically protected edge states at the interface (details in Supplemental Material. C). The transmission of the edge modes in the VTPC is investigated (Fig. 3(b)). It could be seen that the edge modes do not suffer from scattering even at the sharp corner, verifying the topological protection property.

In succession, as shown in Fig. 3(c). disorder is implemented by randomly rotating the Si rods around the center in each unit cell. The random rotation angle follows the uniform distribution of $\theta_i \sim U(-\frac{\theta_d}{2}, \frac{\theta_d}{2})$ for the i-*th* cell. As same as the previous session, the random rotation angle range $\theta_d$ is used to represent the strength of the disorder. It should be mentioned that, the maximum possible value of $\theta_d$ is $(\theta_d)_{\max} = 60°$ to ensure the opposite $C_v$ of two disordered VPCs. The theoretical analysis is similarly conducted using Eq. (3). When $\theta_d = 0°$, $|\lambda_e^P|$ is at its maximum, and decreases as $\theta_d$ increases, while the $\text{sgn}(\lambda_e^P)$ and valley Chern number of two disordered VPCs keeps opposite. Decreased $|\lambda_e^P|$ would reduce the bulk energy gap as well the corresponding edge transmission width. However, with the strongest disorder ($\theta_d = 60°$), $|\lambda_e^P|$ remains greater than 0, so that the disordered photonic crystal keeps topologically protected.

We have simulated the bulk and edge transmission of the disordered VTPCs with $\theta_d = 0°$, 20°, 30°, 40° and 60° (details in Supplemental Material. D). Fig. 3(d) shows a case of $\theta_d = 60°$, it could be seen that even under the strongest disorder, there is still a topologically protected edge mode around the interface. Corresponding transmission spectra are also calculated and plotted in Fig. 3(e). With $\theta_d$ increasing, the width of bulky energy gap and edge transmission would decrease. However, even when $\theta_d = 60°$, there is still maintained topologically protected edge states.

†Yongzhuo Li: liyongzhuo@tsinghua.edu.cn

‡Xue Feng: x-feng@tsinghua.edu.cn

§Yidong Huang: yidonghuang@tsinghua.edu.cn

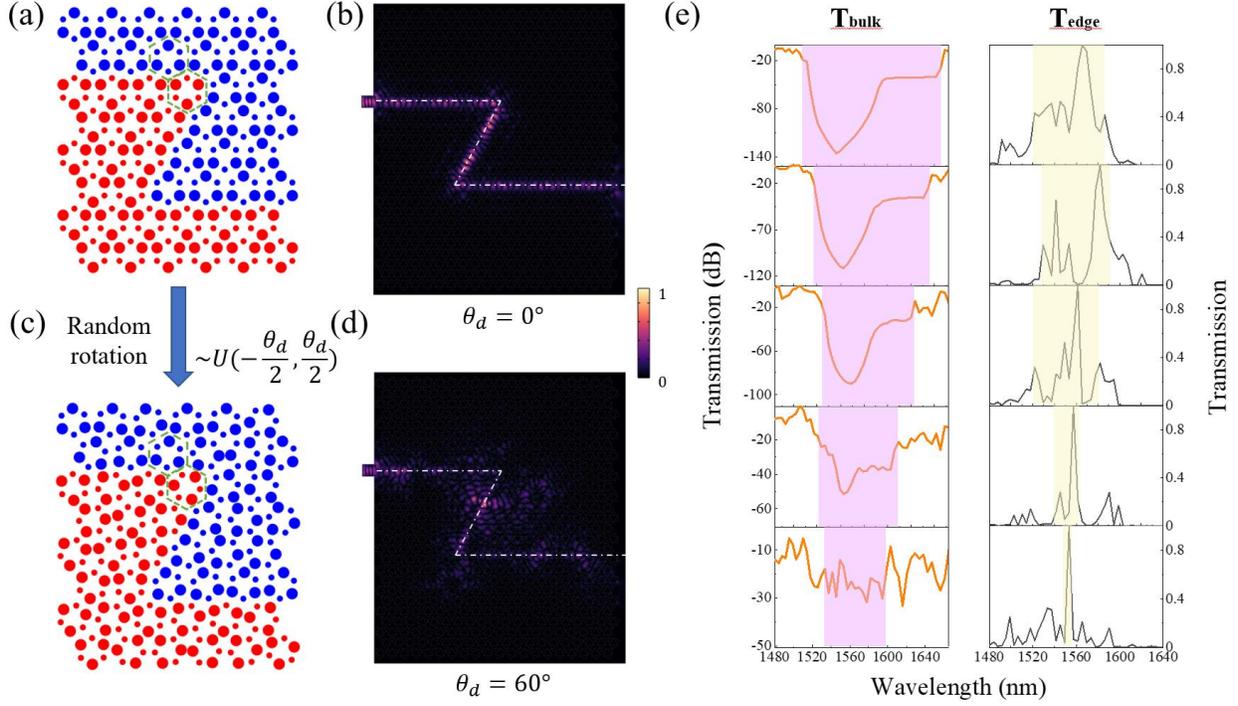

Fig. 3. Schematic of the robust VTPC. (a) Schematic of the stable valley topologically protected photonic crystal composed of two VPCs with different $C_v$, blue (red) represents $C_v = -2$ ($C_v = 2$). (b) Simulated field intensity of $|E_z|$ of the topologically protected edge mode at 1555nm, the interface is marked by white dashed line. (c) Schematic of disordered VTPC, disorder is introduced by randomly rotating the rods in each unit cell. (d) Simulated edge mode field intensity of $|E_z|$ in the disordered VTPC with $\theta_d = 60°$ at 1555nm, the interface is marked by white dashed line. (e) Simulated bulky and edge transmission spectra with $\theta_d = 0°$, 20°, 30°, 40° and 60°, from top to bottom. Bulk energy gap is marked by purple rectangular (defined by transmission < -20 dB), while edge transmission width is marked by yellow rectangular. All the spectra are normalized.

*Experimental demonstrations and results*—Based on the above theoretical and simulation results, we have fabricated the VTAPCs and robust disordered VTPCs. In all samples, two VPCs are arranged to form a Z-shaped interface. Two straight waveguides are also fabricated on both sides of the photonic crystals to excite and collect edge state modes, as the SEM images of Fig. 4(a), 4(d) shown. Correspondingly, vertical grating couplers are also fabricated at the waveguide ends for coupling of input and output light beams. With a tunable laser operating within the wavelength range from 1480 nm to 1640 nm, the edge transmission spectra are measured and characterized (details in Supplemental Material. E and F).

For VTAPCs, samples are demonstrated for each different degree of disorder $\theta_d = 0°$, 15°, 30°, 45° and 60°. Fig. 4(a) shows the SEM images of the $\theta_d = 0°$ case. The edge transmission spectra are plotted in Fig. 4(b), while transmission bandwidths are highlighted by the yellow boxes. With the increase of $\theta_d$, initially there is no edge transmission ($\theta_d = 0°$, 15°), then transmission peak appears ($\theta_d = 30°$), achieves maximum bandwidth ($\theta_d = 45°$) and finally its bandwidth decreases ($\theta_d = 60°$). Normalized $|\lambda_e^P|$, simulated bulk energy gap and measured edge transmission widths are plotted in Fig. 4(c). It could be seen that when $\theta_d = 15°$, $|\lambda_e^P| > 0$, there is no energy gap and edge transmission. The reason is that $\lambda_e^P$ is too small to open the complete bandgap, as mentioned before. When $\theta_d = 30°$, 45°and 60°, the trends of these curves are in good agreement. The topological phase transition has occurred when $\theta_d = 30°$, while such random rotation angle is ~60° in previous works [22,25]. It indicates that our proposed VTAPCs could achieve much lower threshold phase transition, which is attributed to the critical topologically trivial phase. A smaller, and even nearly zero threshold could be further achieved by implementing a VPC with a valley band gap that is equal to or very close to the complete band gap (An example is given in Supplemental Material. G).

†Yongzhuo Li: liyongzhuo@tsinghua.edu.cn

‡Xue Feng: x-feng@tsinghua.edu.cn

§Yidong Huang: yidonghuang@tsinghua.edu.cn

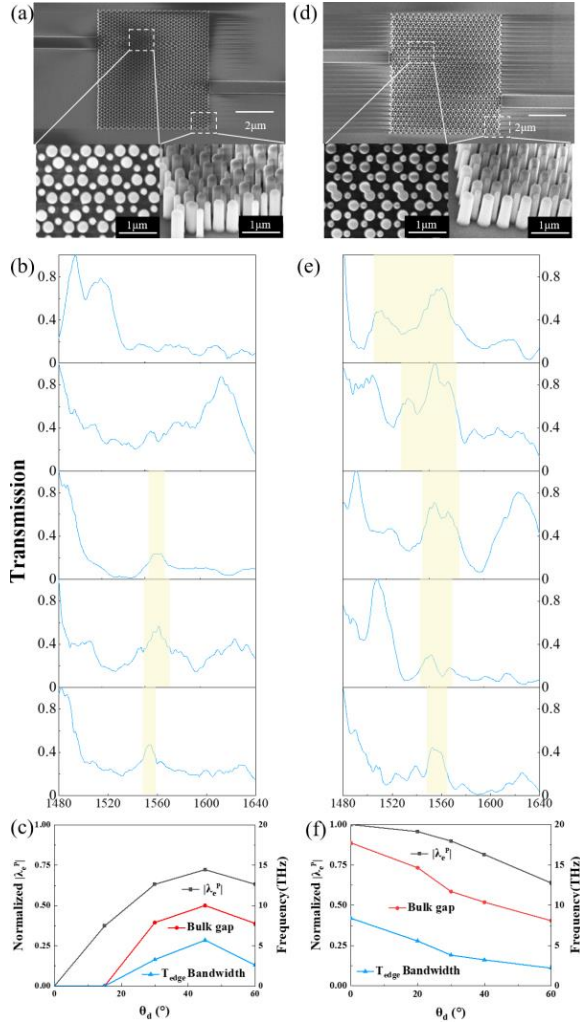

Fig. 4. (a) SEM images of the VTAPC with $\theta_d = 0°$. (b) Measured edge transmission spectra of VTAPCs with different degree of disorder: $\theta_d = 0°$, 15°, 30°, 45° and 60°. From top to bottom, $\theta_d$ increases. (c) Normalized $|\lambda_e^P|$ (black squares), simulated bulk energy gap widths (red circles) and measured edge transmission widths (blue triangles) of VTAPCs with different $\theta_d$. (d) SEM images of the VTPC with $\theta_d = 0°$. (e) Measured edge transmission spectra of disordered VTPCs with $\theta_d = 0°$, 20°, 30°, 40° and 60°. From top to bottom, $\theta_d$ increases. (e) Normalized $|\lambda_e^P|$ (black squares), simulated bulk energy gap widths (red circles) and measured edge transmission widths (blue triangles) of disordered VTPCs with different $\theta_d$. All the spectra are normalized.

†Yongzhuo Li: liyongzhuo@tsinghua.edu.cn

‡Xue Feng: x-feng@tsinghua.edu.cn

§Yidong Huang: yidonghuang@tsinghua.edu.cn

For robust VTPCs, samples are demonstrated for each different degree of disorder $\theta_d = 0°$, 20°, 30°, 40° and 60°. Fig. 4(d) shows the SEM images of the case with $\theta_d = 0°$. Fig. 4(e) summarized the measured edge transmission spectra (yellow boxes highlight the transmission bandwidth). When $\theta_d = 0°$ (disorder-free), a broadband (1504-1570nm) edge state transmission is observed. As $\theta_d$ increasing, the bandwidth of the edge transmission gradually decreases. When $\theta_d = 60°$ (maximum disorder), a narrowband edge transmission still exists within the range of 1548-1564nm. It indicates that the topologically protected edge states are still preserved even in the case of strong disorder. Normalized $|\lambda_e^P|$ (black rods), simulated bulk energy gap (red rods) and measured edge transmission widths (blue rods) are plotted in Fig. 4(f). They are all at their maximum at $\theta_d = 0°$, and decrease with $\theta_d$ increasing. And the edge transmission width is smaller than the bulk transmission width, as the edge state band is located in the middle of the band gap. It could be seen that the trends of these curves are consistent, thus validating the correctness of our proposal.

*Conclusion*—Based on the valley degree of freedom, we have demonstrated a 2D time-reversal symmetric PTAI for the first time, to the best of our knowledge. The phase transition threshold of our work is much lower than previous works, and this is attributed to the critical topologically trivial phase of our proposed structure. With our proposal, near-zero threshold is expected to be achieved, and could also be extended to other topological systems, such as condensed matter and acoustic systems [31–34]. Furthermore, we have also realized an extremely robust valley protected photonic crystal based on stable topological phase. This is beneficial to construct robust PICs, especially quantum PICs [35–38]. Moreover, our work employs silicon platform and operates within the near-IR wavelength range, which provides an outstanding platform to investigate the on-chip integrated topological and disordered systems.

*Acknowledgement*—Funding from the National Key Research and Development Program of China (2023YFB2806703), the National Natural Science Foundation of China (Grant No. U22A6004, 92365210, 62175124) is greatly acknowledged. This work was also supported by Beijing National Research Center for Information Science and Technology (BNRist), Frontier Science Center for Quantum Information, Beijing academy of quantum information science, and Tsinghua University Initiative Scientific Research Program.

---

[1] L. Lu, J. D. Joannopoulos, and M. Soljačić, Topological photonics, Nat. Photonics **8**, 821 (2014).
[2] T. Ozawa et al., Topological photonics, Rev. Mod. Phys. **91**, 015006 (2019).
[3] F. D. M. Haldane and S. Raghu, Possible Realization of Directional Optical Waveguides in Photonic Crystals with Broken Time-Reversal Symmetry, Phys. Rev. Lett. **100**, 013904 (2008).
[4] B. Bahari, A. Ndao, F. Vallini, A. El Amili, Y. Fainman, and B. Kanté, Nonreciprocal lasing in topological cavities of arbitrary geometries, Science **358**, 636 (2017).
[5] G. Harari, M. A. Bandres, Y. Lumer, M. C. Rechtsman, Y. D. Chong, M. Khajavikhan, D. N. Christodoulides, and M. Segev, Topological insulator laser: Theory, Science **359**, eaar4003 (2018).
[6] F. Gao, H. Xue, Z. Yang, K. Lai, Y. Yu, X. Lin, Y. Chong, G. Shvets, and B. Zhang, Topologically protected refraction of robust kink states in valley photonic crystals, Nat. Phys. **14**, 140 (2018).
[7] J.-W. Liu, F.-L. Shi, X.-T. He, G.-J. Tang, W.-J. Chen, X.-D. Chen, and J.-W. Dong, Valley photonic crystals, Adv. Phys. X **6**, 1905546 (2021).
[8] A. B. Khanikaev and G. Shvets, Two-dimensional topological photonics, Nat. Photonics **11**, 763 (2017).
[9] X.-T. He, E.-T. Liang, J.-J. Yuan, H.-Y. Qiu, X.-D. Chen, F.-L. Zhao, and J.-W. Dong, A silicon-on-insulator slab for topological valley transport, Nat. Commun. **10**, 872 (2019).
[10] Y. Yang, Y. Yamagami, X. Yu, P. Pitchappa, J. Webber, B. Zhang, M. Fujita, T. Nagatsuma, and R. Singh, Terahertz topological photonics for on-chip communication, Nat. Photonics **14**, 446 (2020).
[11] M. I. Shalaev, W. Walasik, A. Tsukernik, Y. Xu, and N. M. Litchinitser, Robust topologically protected transport in photonic crystals at telecommunication wavelengths, Nat. Nanotechnol. **14**, 31 (2019).
[12] Y. Gong, S. Wong, A. J. Bennett, D. L. Huffaker, and S. S. Oh, Topological Insulator Laser Using Valley-Hall Photonic Crystals, ACS Photonics **7**, 2089 (2020).
[13] B. Yang, H. Zhang, Q. Shi, T. Wu, Y. Ma, Z. Lv, X. Xiao, R. Dong, X. Yan, and X. Zhang, Details of the topological state transition induced by gradually increased disorder in photonic Chern insulators, Opt. Express **28**, 31487 (2020).
[14] P. Zhou, G.-G. Liu, X. Ren, Y. Yang, H. Xue, L. Bi, L. Deng, Y. Chong, and B. Zhang, Photonic amorphous topological insulator, Light Sci. Appl. **9**, 133 (2020).
[15] C. Liu, W. Gao, B. Yang, and S. Zhang, Disorder-Induced Topological State Transition in Photonic Metamaterials, Phys. Rev. Lett. **119**, 183901 (2017).
[16] P. W. Anderson, Absence of Diffusion in Certain Random Lattices, Phys. Rev. **109**, 1492 (1958).
[17] M. Segev, Y. Silberberg, and D. N. Christodoulides, Anderson localization of light, Nat. Photonics **7**, 197 (2013).
[18] S. Karbasi, R. J. Frazier, K. W. Koch, T. Hawkins, J. Ballato, and A. Mafi, Image transport through a disordered optical fibre mediated by transverse Anderson localization, Nat. Commun. **5**, 3362 (2014).
[19] J. Li, R.-L. Chu, J. K. Jain, and S.-Q. Shen, Topological Anderson Insulator, Phys. Rev. Lett. **102**, 136806 (2009).
[20] C. W. Groth, M. Wimmer, A. R. Akhmerov, J. Tworzydło, and C. W. J. Beenakker, Theory of the Topological Anderson Insulator, Phys. Rev. Lett. **103**, 196805 (2009).
[21] S. Stützer, Y. Plotnik, Y. Lumer, P. Titum, N. H. Lindner, M. Segev, M. C. Rechtsman, and A. Szameit, Photonic topological Anderson insulators, Nature **560**, 461 (2018).
[22] G.-G. Liu et al., Topological Anderson Insulator in Disordered Photonic Crystals, Phys. Rev. Lett. **125**, 133603 (2020).
[23] M. Ren et al., Realization of Gapped and Ungapped Photonic Topological Anderson Insulators, Phys. Rev. Lett. **132**, 066602 (2024).
[24] T. Qu, M. Wang, X. Cheng, X. Cui, R.-Y. Zhang, Z.-Q. Zhang, L. Zhang, J. Chen, and C.

[†] Yongzhuo Li: liyongzhuo@tsinghua.edu.cn

[‡] Xue Feng: x-feng@tsinghua.edu.cn

[§] Yidong Huang: yidonghuang@tsinghua.edu.cn

T. Chan, Topological Photonic Alloy, Phys. Rev. Lett. **132**, 223802 (2024).

[25] X.-D. Chen, Z.-X. Gao, X. Cui, H.-C. Mo, W.-J. Chen, R.-Y. Zhang, C. T. Chan, and J.-W. Dong, Realization of Time-Reversal Invariant Photonic Topological Anderson Insulators, Phys. Rev. Lett. **133**, 133802 (2024).

[26] X. Cui, R.-Y. Zhang, Z.-Q. Zhang, and C. T. Chan, Photonic Z 2 Topological Anderson Insulators, Phys. Rev. Lett. **129**, 043902 (2022).

[27] X.-D. Chen, F.-L. Zhao, M. Chen, and J.-W. Dong, Valley-contrasting physics in all-dielectric photonic crystals: Orbital angular momentum and topological propagation, Phys. Rev. B **96**, 020202 (2017).

[28] T. Ma and G. Shvets, All-Si valley-Hall photonic topological insulator, New J. Phys. **18**, 025012 (2016).

[29] A. B. Khanikaev, S. Hossein Mousavi, W.-K. Tse, M. Kargarian, A. H. MacDonald, and G. Shvets, Photonic topological insulators, Nat. Mater. **12**, 233 (2013).

[30] B. Yan et al., Multifrequency and Multimode Topological Waveguides in a Stampfli-Triangle Photonic Crystal with Large Valley Chern Numbers, Laser Photonics Rev. **18**, 2300686 (2024).

[31] C. P. Orth, T. Sekera, C. Bruder, and T. L. Schmidt, The topological Anderson insulator phase in the Kane-Mele model, Sci. Rep. **6**, 24007 (2016).

[32] Y. Su, Y. Avishai, and X. R. Wang, Topological Anderson insulators in systems without time-reversal symmetry, Phys. Rev. B **93**, 214206 (2016).

[33] H. Liu, B. Xie, H. Wang, W. Liu, Z. Li, H. Cheng, J. Tian, Z. Liu, and S. Chen, Acoustic spin-Chern topological Anderson insulators, Phys. Rev. B **108**, L161410 (2023).

[34] Z.-Q. Zhang, B.-L. Wu, J. Song, and H. Jiang, Topological Anderson insulator in electric circuits, Phys. Rev. B **100**, 184202 (2019).

[35] T. Dai et al., A programmable topological photonic chip, Nat. Mater. **23**, 928 (2024).

[36] C. J. Flower et al., Observation of topological frequency combs, Science **384**, 1356 (2024).

[37] Y. Chen, X.-T. He, Y.-J. Cheng, H.-Y. Qiu, L.-T. Feng, M. Zhang, D.-X. Dai, G.-C. Guo, J.-W. Dong, and X.-F. Ren, Topologically Protected Valley-Dependent Quantum Photonic Circuits, Phys. Rev. Lett. **126**, 230503 (2021).

[38] G.-J. Tang et al., Broadband and fabrication-tolerant 3-dB couplers with topological valley edge modes, Light Sci. Appl. **13**, 166 (2024).

[†] Yongzhuo Li: liyongzhuo@tsinghua.edu.cn

[‡] Xue Feng: x-feng@tsinghua.edu.cn

[§] Yidong Huang: yidonghuang@tsinghua.edu.cn